\documentclass[a4,11pt]{article}
\newdimen\mathindent
\renewcommand{\qquad}{\hspace*{25pt}}
\newcommand{\eref}[1]{(\ref{#1})}

\def\wP{\widehat P}
\def\wPm{$\widehat P$-matrix}
\def\wPms{$\widehat P$-matrices}

\def\prodd#1#2#3{\prod\limits_{#1}^{#2}\lower3pt\hbox{${ }_{#3}$}}

\newcommand{\eqll}[2]{\begin{equation}{#1}\label{#2}\end{equation}}

\def\integer{{\mathchoice
    {\hbox{ $\displaystyle\kern-1mm {\rm Z}\kern-1.1mm {\rm Z}$}}
    {\hbox{ $\textstyle\kern-1mm {\rm Z}\kern-1.1mm {\rm Z}$}}
    {\hbox{$\scriptstyle\kern-1mm {\rm Z}\kern-1.1mm {\rm Z}$}}
    {\hbox{$\scriptscriptstyle\kern-1mm {\rm Z}\kern-1.1mm {\rm Z}$}}}}
\def\natural{{\mathchoice
    {\hbox{ $\displaystyle\kern-1.4mm 1\kern-.7mm {\rm N}$}}
    {\hbox{ $\textstyle\kern-1.4mm 1\kern-.7mm {\rm N}$}}
    {\hbox{$\scriptstyle\kern-1.4mm 1\kern-.7mm {\rm N}$}}
    {\hbox{$\scriptscriptstyle\kern-1.4mm 1\kern-.7mm {\rm N}$}}}}
\def\real{{\mathchoice
    {\hbox{$\displaystyle\kern-.2mm 1\kern-.8mm {\rm R}\kern-.2mm$}}
    {\hbox{$\textstyle\kern-.2mm 1\kern-.8mm {\rm R}\kern-.2mm$}}
    {\hbox{$\scriptstyle\kern-.2mm 1\kern-.8mm {\rm R}\kern-.2mm$}}
    {\hbox{$\scriptscriptstyle \kern-.2mm 1\kern-.8mm {\rm R}\kern-.2mm$}}}}
\def\complex{{\mathchoice
    {\hbox{$\displaystyle\kern-.2mm {\rm C}\kern-1.5mm\raise.2mm
                   \hbox{\vrule height6pt}\kern1.3mm$}}
    {\hbox{$\textstyle\kern-.2mm {\rm C}\kern-1.5mm\raise.3mm
                   \hbox{\vrule height6pt}\kern1.3mm$}}
    {\hbox{$\scriptstyle\kern-.2mm{\rm C}\kern-1.5mm\raise.2mm
                   \hbox{\vrule height3pt}\kern1.3mm$}}
    {\hbox{$\scriptscriptstyle\kern-.2mm{\rm C}\kern-1.5mm\raise.2mm
                  \hbox{\vrule height2pt}\kern1.3mm$}}}}
\def\summ#1#2#3{\sum\limits_{#1}^{#2}\lower3pt\hbox{${ }_{#3}$}}

\def\prodd#1#2#3{\prod\limits_{#1}^{#2}\lower3pt\hbox{${ }_{#3}$}}

\begin{document}

\title{The Orbit Space Approach to the Theory of Phase Transitions:
The Non-Coregular Case}

\author{ G. Valente \small
\\
\small Dipartimento di Fisica Universit\`{a} di Padova \\ \small
and INFN, Sezione di Padova,\\ \small via Marzolo 8, I-35131
Padova, Italy}

\maketitle
\begin{abstract}
We consider the problem of the determination of the isotropy
classes of the orbit spaces of all the real linear groups, with
three independent basic invariants satisfying only one independent
relation. The results are obtained in the $\widehat P$-matrix
approach solving a universal differential equation ({\em master
equation}) which involves as free parameters only the degrees
$d_a$ of the invariants. We begin with some remarks which show how
the $\widehat{P}$-matrix approach may be relevant in physical
contexts where the study of invariant functions is important, like
in the analysis of phase spaces and structural phase transitions
(Landau's theory).
\end{abstract}

\section{The Orbit Space (OS) Approach}
Invariant functions under the transformations of a Compact Linear
Group (CLG) acting in $\real^n$ can be expressed in terms of
functions defined in the OS of the group, i.e.\ as functions of a
finite set of basic invariant polynomials $p(x) \equiv (p_1(x
),\dots ,p_q(x))$, $x\in\real^n$, which may be chosen to form a
Minimal Integrity Basis (MIB) for the group $G$. Such an
observation, originally due to Gufan (1971), simplifies the
determination of the patterns of spontaneous symmetry breaking
(SSB) in theories in which the ground state is determined by the
minimum of an invariant potential $V(x)$. When $p$ ranges in the
domain spanned by $p(x),\ x\in \real^n$, the function $\widehat
V(p)$ has the same range as $V(x)$, but is not plagued by the same
degeneracies. A correct exploitation of this idea required,
however, the determination of the ranges of the functions
$p_i(x)$, a problem which was completely solved only using the
powerful tools of geometric invariant theory. An excellent review
with motivations and references was written by G. Sartori
\cite{683}. He discovered a simple recipe allowing to determine
the structure of the OS of any CLG (the \wPm\/ approach). The OSs
admit an analytical image in terms of connected semi-algebraic
varieties, whose defining relations can be expressed from
(semi)positivity conditions of matrices $\widehat{P}(p)$, which
are constructed only through the knowledge of a MIB for the group.
In solid state physics, the OS approach was applied to the Landau
theory of phase transitions, but it was mainly reduced to a naive
numerical technique. Since the free energy expansion is commonly
considered up to the 6th degree in the order parameter, the OS
 was realized through a ``projection'' onto the subspace of $\real^q$
corresponding to invariants $p_a(x)$ such that $\mbox{\rm deg}\, p_a(x)
\leq 6$.  On the contary, using the \wPm-approach,
 it is possible to get an exact, complete
analytical determination of the primary stratification of the OS,
as it was shown for the ``classical'' example of $\mbox{\rm Ba Ti
O}_{3}$ transitions \cite{687}. Nevertheless the complete MIB's
are at disposal only for finite groups generated by reflections,
and for simple Lie groups \cite{687,ult}. A way to obtain the
matrices $\wP(p)$ generated by CLG's, bypassing the problem, is to
use an axiomatic approach, which is based on the notion of MIB
transformations \cite{681}. An equivalence relation is defined on
the set of \wPm\/, consequently on the set of OSs. Thus,
universality properties of the schemes of SSB are pointed out;
here, we shall examine
 the non-coregular case.

\section{Non-Coregular Compact Linear Groups}
Given a MIB $p(x)\equiv \left(p_1(x),\ldots,p_q(x)\right)$,
 if polynomial functions $\widehat{F}$ exist such
that $\widehat{F}\left(p_1(x),\ldots,p_q(x)\right) \equiv 0 $, then $G$
is said to be {\em non-coregular}.

\noindent {\em Examples:} Consider the 2-dimensional point groups $C_n$,
$n \geq 2$ and the 3-dimensional point groups $Y$, $O$, $T_h$, $T$.

Let $\{\widehat F_A(p)\}_{1\le A\le K}$
be a complete set of basic homogeneous relations among the elements of a
MIB.
The polynomials $\widehat F_A(p)$ can be chosen to be
$w$-ho\-mo\-ge\-neo\-us and irreducible on $\complex$.  The associated
equations
$\widehat F_A(p)\,=\,0$, define an irreducible algebraic variety
in $\real^q$
(and in $\complex^{q}$ for $ p \in \complex^{q}$),
the
{\em variety $\cal Z$ of the relations} among the elements of the MIB.

The variety $\cal Z$ has a singularity in $p = 0$.
In fact, for all $A$, $\widehat F_A (p)$
cannot be solved polynomially
with respect to anyone of the basic invariants $p_a$.
The absence of linear terms in any $p_a$ implies the vanishing of
$\widehat{F}_{A}(0)$ and
$\partial\widehat{F}_{A}(0)$ for all $A=1,\dots, K$.
For $k=\mbox{dim}({\cal Z})$, the couple $( q ,k)$ will define the {\it regularity
type} (called $r$-type) of $G$.

\section{Characterizing the matrices $\wP(p)$}
The \wPms\ associated to CLG's
will be characterized through a set of structural properties
that can be put in the form of differential relations.

Therefore,
forgetting altogether the original definition of the
matrices $\wP(p)$, we shall be able to
determine the \wPms\
as solutions of a system of differential
equations obeying convenient initial conditions.

Although we cannot enter the details of the generalization of the
coregular approach \cite{683}, we shall point out the new {\em
initial conditions} which fit the non-coregular case
\cite{noncor}. We shall show the striking results obtained for
groups of r-type $(q,q-1)$ in the simplest case of MIBs having
cardinality $q=3$.

Let $\sigma$ be a general primary stratum of $\overline{{\cal
S}}\equiv p(\real^n)$, and  ${\cal I}({\sigma})$ the ideal formed
by all the polynomials in $p \in \real^{q}$ vanishing on
${\sigma}$.  Every $\hat{f}(p)\in {\cal I}({\sigma})$ defines in
$\real^{n}$ an invariant polynomial function $f(x) =
\hat{f}(p(x))$, and $f(x)  \,=\, 0 $ for all $x \in \Sigma_{f}=p^{
-1}({\sigma})$. As in the coregular case, the gradient $\partial
f(x)$ is obviously orthogonal to $\Sigma_f$ at every
$x\in\Sigma_f$, but, it must also be tangent to $\Sigma_{f}$ since
$f(x)$ is a G--invariant function \cite{681}. Consequently, it has
to vanish on $\Sigma_{f}$:
\begin{equation} \label{A1}
0\,=\,\partial f(x) \,=\, \left.\summ 1qb\partial_b \hat{f}(p)
\, \partial p_{b}(x) \right|_{p=p(x)}
\,, \hspace{2em} \forall x \in \Sigma_{f} \,.
\end{equation}
By taking the scalar product of (\ref{A1}) with $\partial p_{a}(x)$, we
end up with the following {\em boundary conditions}:

\begin{equation} \label{A2}
\summ 1qb \widehat{P}_{ab}(p) \, \partial_b \hat{f}(p) \in
{\cal I}({\sigma}), \hspace{2em} \forall \hat{f} \in {\cal I}
({\sigma}) \,\, \mbox{and}
\,\, \forall {\sigma} \subseteq \overline{{\cal S}}\,.
\end{equation}
Relation (\ref{A2}) can be re-proposed in the form of a differential relation
involving only polynomial functions of $p$.
According to the Hilbert basis theorem, the ideal ${\cal
I}(\sigma)$ is finitely generated.
Let $\{ f^{(1)}(p), f^{(2)}(p), \ldots, f^{(m)}(p) \}$ be a $w$-homogeneous
basis for ${\cal I}({\sigma})$, then \eref{A2} is equivalent to the
following relations:

\begin{equation} \label{A3}
\summ 1qb \widehat{P}_{ab}(p) \, \partial_b f^{(r)}(p)
 \,=\, \summ 1ms \lambda^{(rs)}_a(p) \, f^{(s)}(p)\,,
\qquad a=1,\ldots,q\,; \;\; r=1,\ldots,m \,,
\end{equation}
where the $\lambda^{(rs)}$'s are $w$-homogeneous polynomial functions of
$p$.

If, in particular,  $\sigma$ is a $(q-1)$-dimensional primary
stratum, the ideal ${\cal I}(\sigma)$ has a unique {\it irreducible}
generator, $a(p)$, and \eref{A3} reduces to the simpler form

\begin{equation} \label{mas}
\summ 1qb \widehat P_{ab}(p) \, \partial_b a(p) \,=\, \lambda_a(p)\, a(p)\,,
\qquad a=1, \dots, q\,.
\end{equation}
Equation (\ref{mas}) will be quoted as {\em master relation}.
It is valid  for the principal stratum of the OS of a group of
$r$-type $(q,q-1)$.

The structure of  (\ref{mas}) has been completely analyzed
\cite{682}. We just recall that, like in the coregular case,
$a(p)$ is a polynomial factor of $\det \wP(p)$; besides, there
exist particular MIB's, which we shall call {\it $A$-bases}, in
which the master relation takes the following {\it canonical}
form: \eqll{\summ 1qb \widehat P_{ab}(p)\partial_b A(p) \,=\,
2\delta_{aq} w(A) A(p),\qquad a=1,\dots ,q\,.}{can}

But  there
are further properties which are {\em peculiar to the coregular case}.
They play the role of {\em initial conditions}, because they are so
restrictive to select among the solutions of the canonical equation just
the ones which may correspond to (images) of OSs of
coregular CLGs {\em (allowable solutions)}.

\begin{itemize}

\item[i)] Consider the hyperplane
$\Pi=\{p\equiv(p_1,\ldots,p_q) \, |\, p_q=1\}$ of $\real^q$.
The restriction $A(p)\big |_{\Pi}$, of $A(p)$ to $\Pi$
has a unique local non
degenerate maximum lying at $p^{(0)}=(0,\ldots,0,1)$.
\item[ii)] The matrix $\widehat P(p^{(0)})$ is block diagonal
and, for  {\it standard} $A$-bases, it is diagonal:
$\widehat P_{ab}(p^{(0)})\,=\,d_ad_b\delta_{ab}$, $(a,b=1, \dots,
q)$.

\end{itemize}

\subsection{\label{AddNC}Second order boundary conditions}
The \wPms\ generated by non-coregular groups do not satisfy the set of
``initial conditions'' specified above, but the
presence of relations connecting the elements of any MIB gives rise to
interesting constraints.

Let us consider a compact non-coregular group $G$, whose OS
is realized as a
semi-algebraic subset $\overline{{\cal S}}$ of the variety ${\cal Z}$ of
the relations.
Let ${\cal I}({\cal Z})$ be the ideal of the polynomial
functions of $p$ vanishing on $\cal Z$.  Any polynomial $\widehat{F}(p) \in
{\cal I}({\cal Z})$ defines an identity in $\real^{n}$:

\begin{equation} \label{BB1}
F(x) \,=\, \widehat{F}(p(x)) \,=\, 0 \,,\quad \forall x \in \real^n\,,
\end{equation}
which, after differentiating twice with respect to $x_i$
and summing over $i$, gives rise to
the following condition, valid $\forall x \in \real^{n}$:

\begin{equation} \label{BB3}
 \summ 1ni \left \{ \summ 1q{a,b}
\partial_{a b} \widehat{F}
(p(x))
\, \partial_{i} p_{b}(x) \,
   \partial_{i} p_{a}(x) +
\summ 1qa
\partial_{a} \widehat{F}
(p(x))
\, \partial^{2}_{i} p_{a}(x)
\right \}  \,=\, 0 \,.
\end{equation}

Since $G$ is a matrix subgroup of $\mbox{O}_{n}(\real )$,
the $n$-dimensional
Laplacian of any invariant polynomial function of $x$ is a
$G$-invariant polynomial.  Thus Hilbert's theorem ensures
the existence of a
set $\{\hat l_{a}(p)\}_{a=1, 2, \ldots, q}$ of
$w$-homogeneous
polynomials in $p\in \real^q$ with weights
$w(\hat l_{a}) = d_{a} - 2$,
such that:

\begin{equation}\label{BB3'}
l_a(x)\ =\  \summ 1ni \partial^{2}_{i} p_{a}(x) \,=\,
\hat l_{a}(p(x))\,, \qquad a=1, 2, \dots, q \,.
\end{equation}

The identity expressed in (\ref{BB3}) induces, through the orbit map, the
following polynomial relation, valid for all $p\in \overline{{\cal S}}
\,=\,p(\real^{n})$,
and consequently, for all $p \in  {\cal Z}$,
as $p(\real^{n})$ is a semi-algebraic subset of $\cal Z$ of the same
dimension as $\cal Z$:

\begin{equation} \label{FE1}
\summ 1q{a,b} \widehat{P}_{ab}(p)\;\partial_{a}\partial_{b} \widehat{F}(p)
  +
 \summ 1qa \hat{l}_{a}(p)\; \partial_{a} \widehat{F}(p)
     \,=\, 0 \,, \qquad   p \in {\cal Z}\;.
\end{equation}

We stress that, owing to the convention $p_{q}=\sum_{j=1}^{n}
x_{j}^{2}$, we get $l_{q} \,=\, 2 n$. Even if in our approach the
explicit form of the polynomials of a MIB is not specified, we
have a clue of the power of relation (\ref{FE1}), since it is
linked with the effective group action. The parameter $n$
interprets the role of the dimension of the real space in which
$G$ acts. Therefore (\ref{FE1}) will be considered as a sort of
{\em second order boundary condition} in which $l_{q} \,=\, 2
n\,\ge 4$ \cite{noncor}.

If the group $G$ is of r-type $(q,q-1)$, the ideal ${\cal I}({\cal
Z})$ has a unique generator $\widehat F(p)$, which fulfils the
canonical equation (\ref{can}). It may be proved that (\ref{FE1})
holds identically $\forall p \in \real^q$, essentially for weight
reasons \cite{noncor}. The solution procedure consists in
expressing all the polynomials involved in (\ref{can}) and
(\ref{FE1}) in the most general form in agreement with their
structural properties. For groups $G$ of r-type $(3,2)$ we get a
complicated system of coupled algebro-differential equations. Some
simplifications are obtained replacing (\ref{FE1}) with a form in
which it appears only the first order derivative of
$\widehat{F}(p)$. The $\partial_3 \widehat{F}(p)$ term can be
eliminated using  the w-homogeneity conditions on
$\widehat{F}(p)$.

\subsection{Allowable $\widehat{P}$-matrices: the non-coregular case}
A {\em proper solution}
is  a couple $(\widehat{P}(p),\widehat{F}(p))$, where $\widehat{F}(p)$
is the factor of $\mbox{\rm det} \widehat{P}(p)$ which defines the
variety ${\cal Z}$.
Consider the semialgebraic variety
${\cal R}^{(\geq)} = \{p \in {\cal Z} \,|\, \wP(p)
\geq 0 \}$.
In our approach for non-coregular groups,
${\cal R}^{(\geq)}$ and its algebraic stratification
may {\em potentially} be identified with (the image of) the OS of an
actual group only if the following conditions hold:
\begin{itemize}
\item[i)] On $\cal Z\;$,
$\mbox{{\rm rank}}\, (\wP(p))\le k$ and the set ${\cal R} = \{p \in {\cal Z}
\,|\, \wP(p) \geq 0\,,\, \mbox{{\rm rank}}(\wP(p)) = k \}$ is
$k$-dimensional and connected; its closure
$\overline {\cal R} \equiv {\cal R}^{(\geq)}$.
\item[ii)] $\wP(p)$ satisfies the boundary conditions  {\rm (\ref{A2})},
for each primary stratum ${\sigma_{i}}^{(\alpha)}$ of
$\overline{\cal R}$, and the second order boundary condition
{\rm (\ref{FE1})} for each $\widehat{F} (p) \in {\cal I}({\cal Z})$.
\end{itemize}
If that is the case, we speak of {\em allowable} non-coregular solution
$(\widehat{P}(p),\widehat{F}(p))$.
\section{The results}
We proved that Eq.~\ref{can} with Eq.~\ref{FE1} as initial
condition admits only 3 families of proper solutions
\cite{noncor}.

\noindent
{\bf Solution family S1:} It is found in correspondence with the degrees
$ d_1=k(1+2m)\,, \;\; d_2=2k\,,$ for $k,m \in
\natural_{*}$. It depends on the parameter $\epsilon = \pm 1$. For
$\epsilon=+1$, the set ${\cal R}^{(\geq)}$ is not connected, neither the
boundary conditions are satisfied at the peripherical strata. For
$\epsilon=-1$, the set ${\cal R}^{(\geq)} \cap \Pi$ is 0-dimensional.
Thus, this family does not determine allowable \wPms.

\noindent {\bf Solution family S2:} The degrees are $ d_1=6k\,,
\;\; d_2=4k\,,$ for $k\in \natural_{*}$. It is a one-parameter
collection of distinct classes of equivalent \wPms\/, which
defines an allowable solution only for the value $z=0$ of the
parameter. Nevertheless, the generator of ${\cal Z}$ is
$\widehat{F}(p) ={p_1}^2 + {p_2}^{3}$. Thus, the existence of a
generating group can be excluded on the basis of a general
proposition \cite{noncor}.

\noindent {\bf Solution family S3:} Among the 4 distinct classes
of proper solutions, we get only 1 class of allowable \wPms\/
which exactly coincides with the coregular solution of r-type
$(3,3)$ for the class I(1,1) \cite{682}. Generating coregular
groups of the first element of the family are, for instance, the
linear groups $\mbox{SO}(n, \real)$ acting in $\real^{n} \oplus
\real^{n}$ for $n \geq 3$.

Thus we can assert that coregular and non-coregular groups may share the
same \wPm. The non-coregular solution for degrees
$d_1=d_2=k\geq 2,\; k\in \natural$ corresponds for instance to
the action of a $2$-dimensional representation of
the point group $C_{n}$, $n \geq 2$, that
is the cyclic group of rotations about an axis of the
$n$-th order.
The variety ${\cal Z}$ is determined by the relation
$p_1(x)^2 \,=\, p_3(x)^n - p_2(x)^2$; the image of the OS
$\overline{{\cal S}}\cap \Pi$ is the unit
circumference.

Finally, if the action of $G$ is restricted to the unit sphere $(p_3=1)$,
all the orbit spaces of groups of r-type $(3,2)$ turn out to be
isomorphic.

\section*{Acknowledgments}
My special thanks to Prof. G. Sartori for reading the manuscript and
for all the suggestions and helpful discussions during my Ph.D. studies.


\begin{thebibliography}{10}

\bibitem{683}  G. Sartori,
{\em La Rivista del Nuovo Cimento} {\bf 14}, 1 (1991).

\bibitem{687}
G. Sartori and G. Valente, {\em
J. Phys. A: Math. Gen.} {\bf 29}, 193 (1996).

\bibitem{ult} G. Sartori and V. Talamini, {\em J. Math. Phys.} {\bf 39},
2367 (1998).

\bibitem{681} G. Sartori, {\em Mod. Phys. Lett.} {\bf A 4}, 91 (1989).

\bibitem{noncor}
G. Sartori and G. Valente, {\em Quaderni del CNR} {\bf 54}, 204
(1998).

\bibitem{682} G. Sartori and V. Talamini,
{\em Commun. Math. Phys.}, {\bf 139}, 559 (1991).





\end{thebibliography}
\end{document}